\documentclass[twocolumn]{revtex4}
\usepackage{graphicx,amsmath,amssymb,mathrsfs}
\usepackage{epsfig}

\begin{document}

\title{Transition from lognormal to $\chi^2$-superstatistics for financial time series}

\author{Dan Xu and Christian Beck}

\affiliation{Queen Mary University of London, School of Mathematical Sciences, Mile End Road, London E1 4NS, UK}

\begin{abstract}
Share price returns on different time scales can be well modeled by a superstatistical dynamics.
Here we provide an investigation which type of superstatistics is most suitable to
properly describe share price dynamics on various time scales. It is shown that while
$\chi^2$-superstatistics works well on a time scale of days, on
a  much smaller time scale of minutes
the price changes are better described by lognormal superstatistics. The system dynamics thus exhibits
a transition from lognormal to $\chi^2$ superstatistics as a function of time scale. We
discuss a more general model interpolating between both statistics which fits the observed data very well. We also
present results on correlation functions of the extracted superstatistical volatility parameter, which exhibits
exponential decay for returns on large time scales, whereas for returns on small time scales there are
long-range correlations and power-law decay.

\end{abstract}
\maketitle

\section{Introduction}

   Many well established concepts in mathematical finance (such as the Black-Scholes model) are based on the assumption that an index or a stock price follows a geometric Brownian motion, and as consequence the log returns of these processes are Gaussian distributed. But nowadays it is well known that the log returns of realistic stock prices are typically non-Gaussian with fat tails \cite{ref1}--\cite{tsallisbook}. Such behaviour can be well captured by superstatistical models \cite{ref2}--\cite{thurner}. The basic idea of this method borrowed from nonequilibrium statistical mechanics is to regard the time series as a superposition of local Gaussian processes weighted with a process of a slowly changing variance parameter, often called \(\beta\). This approach has been applied to many areas of complex systems research, including turbulence, high energy scattering processes,
   heterogenous nonequilibrium systems, and econophysics (see e.g. \cite{ref9} for a short review).
    In finance early applications of the superstatistics concept were worked out
    by Duarte Queiros et al. \cite{queiros1,queiros2} and Ausloos et al. \cite{ausloos}.
    Van der Straeten and Beck \cite{ref3} analysed daily closing prices of the Dow Jones Industrial Average index (DJI) and the S\(\and\)P 500 index. They verified that both log-normal superstatistcs and \(\chi^2\) superstatistics result in good approximations. Biro and Rosenfeld \cite{ref4} also studied the data sets of the Dow Jones index and verified that the distribution of log returns is well fitted by a Tsallis distribution. Katz and Li Tian \cite{ref1} showed that the probability distributions of daily leverage returns of 520 North American industrial companies during the 2006-2012 financial crisis comply with the $q$-Gaussian distribution which can be generated by \(\chi^2\) superstatistics. They also verified in \cite{ref2} that the Tsallis entropic parameter $q$ obtained by direct fitting to $q$-Gaussians coincides with the \(q\) obtained from the shape parameters of the \(\chi^2\) distribution fitted to the histogram of the volatility of the returns. Gerig, Vicente and Fuentes \cite{ref6} consider a similar model that indicates that the volatility of intra day returns is well described by the \(\chi^2\) distribution, see also \cite{ref7} for related work in this direction.

    In this paper, we will carefully analyse for various data sets of historical share prices which type of superstatistics is best suited to model the dynamics. While Tsallis statistics ($=q$-statistics) is known to be equivalent to $\chi^2$ superstatistics \cite{beck-cohen, tsallisbook}, there are other types of superstatistics, such as lognormal superstatistics and inverse $\chi^2$ superstatistics \cite{ref8}, which are known to be different
    from $q$-statistics (though all these different statistics generate similar
      distributions if the variance of the fluctuations in $\beta$ is small \cite{beck-cohen}). We show that in our analysis $\chi^2$-superstatistics appears best suitable to describe the daily price changes, whereas on much smaller time scales of minutes lognormal superstatistics seems preferable. We analyse the relevant time scale of the changes in the superstatistical parameter $\beta$ and present results for the decay of correlations in $\beta$. For small return time scales, correlation functions
    exhibit power law decay and there are long memory effects. In the final section, we develop a synthetic stochastic model that fits the data well. This is kind of a hybrid model interpolating between lognormal and $\chi^2$-superstatistics.

    This paper is organized as follows. In section II we look at share price returns on large (daily) time scales.
    In section III we do a similar analysis on small (minute) time scales. In section IV we investigate
    correlations of the superstatistical volatility parameter on both time scales. In section V the hybrid model
    is introduced. Our final concluding remarks are given in section VI.

\section{Superstatistics of log-returns of share prices on a large time scale}

Non-equilibrium system dynamics can often  be regarded as as superposition of a local equilibrium dynamics and a slowly fluctuating process of some variance variable \(\beta\) \cite{beck-cohen}. These types of `superstatistical' nonequilibrium models
are also useful for financial time series \cite{queiros1, queiros2}.
In this article, the empirical data we use as an example
is the historical stock prices of Alcoa Inc(AA), which is an American company that engages in the production and management of primary aluminium, fabricated aluminium and alumina. We have looked at shares
of many other companies as well (see Tab.~1 in section IV), with similar results.
Our data set covers the period January 1998 to May 2013. We study the log return \(R_i\) denoted by
\begin{equation}
 R_i=\log{\left(\frac{S_{i+1}}{S_{i}}\right)}
 \label{logreturn}
\end{equation}
where \(i=0,1,2,...,N\); \(S_i\) and \(S_{i+1}\) are two successive daily closing prices. We consider the normalised log returns
\begin{equation}
 u_i=\frac{R_i-\left<R\right>}{\sqrt{\left<R^2\right> - \left<R\right>^2}}
\end{equation}
which have been rescaled to have variance 1. The symbol $\langle \cdots \rangle$ denotes the
long-time average.

From the simplest superstatistics model point of view, the entire time series of stock prices can be divided into \(n\) smaller time slices \(T\). We call \(T\) optimal window size. Within each \(T\), the financial volatility \(\beta\) is temporarily constant and the log return of the stock price is Gaussian distributed. \(\beta\) has some probability distribution \(f\left(\beta\right)\)
to take a particular value in a given slice. The conditional probability \(p(u|\beta)\) is

\begin{equation}
p(u|\beta)=\sqrt{\frac{\beta}{2\pi}} \exp{\left(-\frac{1}{2}\beta u^2\right)}
\end{equation}
and the marginal probability distribution of \(u\) for long time observation is the average over local Gaussians weighted with the probability density \(f(\beta)\)

\begin{equation}
p(u)=\int p(u|\beta)f(\beta)d\beta .
\label{eq4}
\end{equation}
The integration over \(\beta\) yields non-Gaussian behaviour with fat tails.

We now describe our technique to obtain the optimal window size \(T\) for a given time series. Firstly we split the time series into
\begin{equation}
n=
 \lfloor
 \frac{N}{\Delta t}
 \rfloor
\end{equation}
equal intervals,  {where $\lfloor \; \rfloor$ denotes the floor function and
\(\Delta t \) is the dimensionless window size, i.e. the number of data points in a given window.
$N$ is the total number of data points of the entire time series.} Generally the kurtosis of a random variable $u$ is
defined as
\begin{equation}
\kappa  =\frac{
\langle u^4 \rangle}{
\langle u^2 \rangle^2}
\end{equation}
and it is equal to 3 for a Gaussian distribution of arbitrary variance.
For a given window size \(\Delta t\), the kurtosis in the \(j\)th window is given by

\begin{equation}
\kappa_{\Delta t} {(j)}=\frac{
\frac{1}{\Delta t}{\sum_{i=(j-1)\Delta t+1}^{j\Delta t} u_{i}^{4}}}{
\left(\frac{1}{\Delta t}{\sum_{i=(j-1)\Delta t+1}^{j\Delta t} u_{i}^{2}}\right)^2},
\end{equation}
where \(j=1,2,...,n.\) When we have all the values of kurtosis for all windows, we can calculate an average kurtosis of
the \(n\) windows as
\begin{equation}
\bar{\kappa}_{\Delta t} = \frac{1}{n}\sum_{j=1}^{n} \kappa_{\Delta t} {(j)}.
\end{equation}

The aim is to achieve an optimum window size such that for a given data set the
distribution in each window is as close as possible to a Gaussian, but with varying
variance.
For this purpose the optimal window size \(T\) should satisfy the condition
\begin{equation}
\bar{\kappa}_{\Delta t} = 3.
\label{eq9}
\end{equation}

\begin{figure}
\includegraphics[scale=0.5]{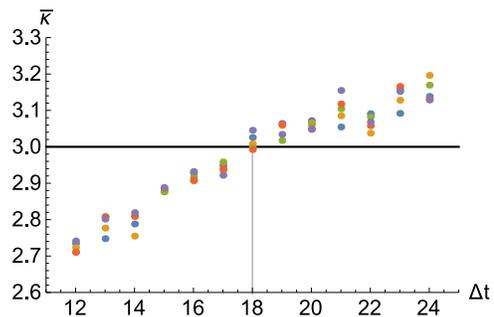}
\caption{{Determination of the optimal window size
for the Alcoa share price data. The intersection with the line kurtosis \(\bar{\kappa}=3\) yields \(T = 18\pm 0.5 \).
The various values of $\bar{\kappa}$ for a given $\Delta t$ (indicated by different colors in the online version)
are obtained for different translational shifts of the sliding windows. The
scattering of the data can be used to estimate the standard deviation as $\delta \bar{\kappa} \sim 0.03$.}}
\label{fig1}
\end{figure}
Fig.~\ref{fig1} shows how the average kurtosis changes with the window size.
 We obtain from condition (\ref{eq9}) the optimal window size $18\pm 0.5$ for this example.
  The result makes financial sense. 18 trading days correspond to a time scale of about 3-4 weeks.
  It is a typical time scale where market volatility changes, due to events such
  as changes in the confidence in the future economic development, anticipated interest changes, and so on.
  See also \cite{camilleri} for related work.

With the given optimal window size, we can now calculate the local volatility parameter $\beta$ in each
time interval as
\begin{equation}
 \beta_k=\frac{1}{\frac{1}{T-1}\sum_{i=(k-1)T+1}^{kT} (u_{i}- \bar{u_i})^2}
\end{equation}
where \(k=1,2,...,n\). Note that the variance of \(u\) in each window is \(\beta^{-1}\).
One can then plot a histogram of the \(\beta_k\) and fit it with some suitable model distribution.

Here we will consider three distributions to be compared with our experimental distribution of \(\beta\),
which were previously advocated in \cite{ref8}. The first one is the \(\chi^2\) -distribution for which \(f(\beta)\) is given by
\begin{equation}
 f_1\left(\beta\right)=\frac{1}{\Gamma{\left(\frac{d_1}{2}\right)}}\left(\frac{d_1}{2\beta_0}\right)^{d_1/2}\beta^{d_1/2-1}e^{-d_1\beta/2\beta_0}.
 \label{eq11}
\end{equation}
The second one is the inverse \(\chi^2\)-distribution where
\begin{equation}
 f_2\left(\beta\right)=\frac{\beta_0}{\Gamma{\left(\frac{d_2}{2}\right)}}\left(\frac{d_2\beta_0}{2}\right)^{d_2/2}\beta^{-d_2/2-2}e^{-d_2\beta_0/2\beta}.
 \label{eq12}
\end{equation}
The third distribution that will be tested is the log-normal distribution for which the probability density function is given by
\begin{equation}
 f_3\left(\beta\right)=\frac{1}{\sqrt{2\pi}s\beta}\exp{\left(\frac{-\left(\ln{\beta}-{\mu}\right)^2}{2s^2}\right)}
\label{eq13}
\end{equation}
where \begin{equation}
\mu=\ln{\beta_0-\frac{s^2}{2}}.
\label{eq14}
\end{equation}
The \(\beta_0\) in Eq. (\ref{eq14}), (\ref{eq11}), (\ref{eq12}) is the mean value of \(\beta\), given by
\begin{equation}
\beta_0=\langle \beta \rangle = \frac{1}{n} \sum_{k=1}^{n} \beta_k,
\end{equation}
and $d_1,d_2,s$ are parameters. Lognormal superstatistics often occurs for
complex systems described by a cascading dynamics \cite{ref10},
whereas $\chi^2$ and inverse $\chi^2$ superstatistics are more common for additive
degrees of freedom contributing to a fluctuating temperature or inverse temperature \cite{ref8}.

We have fitted our experimental histograms \(f(\beta)\) with the above distributions. Given \(\beta_0\), we vary \(d_1, d_2\) and \(s\) of Eq. (\ref{eq11}), (\ref{eq12}), (\ref{eq13}) in order to obtain the optimum fit to our observed \(f(\beta)\).
    \begin{figure}
        \includegraphics[scale=0.5]{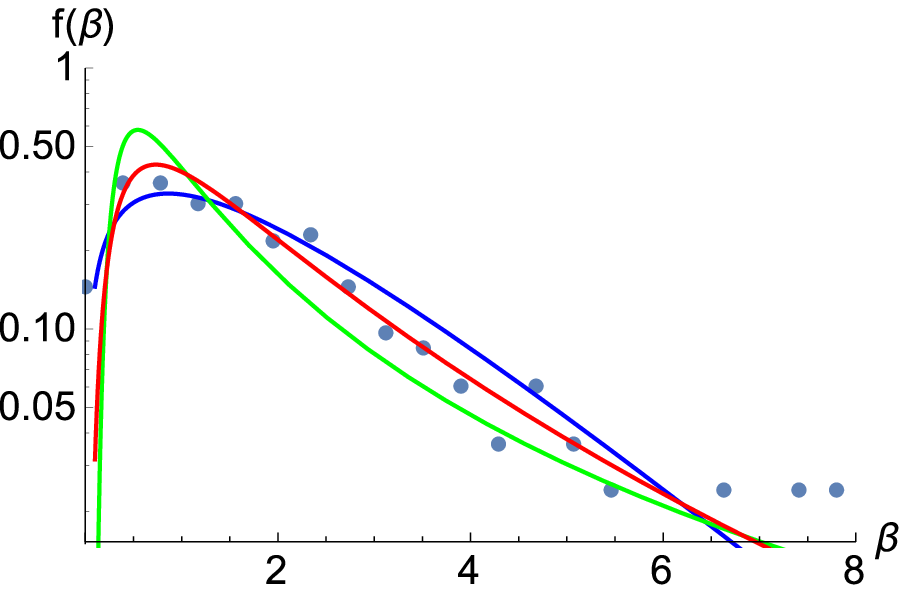}
        \label{fig2a}
        \includegraphics[scale=0.5]{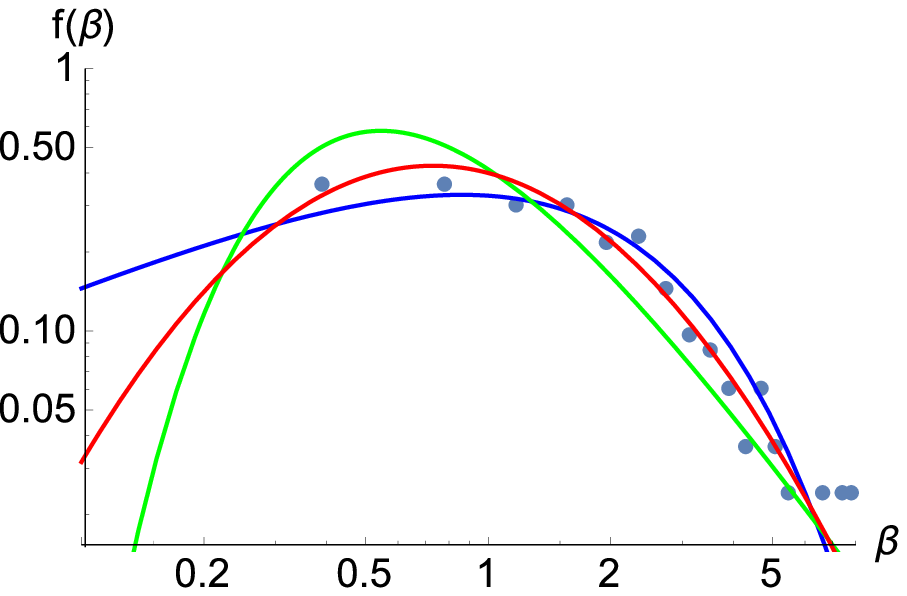}
        \label{fig2b}
    \caption{Best possible fits that can be achieved for the distribution of the volatility \(\beta\) of Alcoa shares (plotted by dots), in a log-linear (top) and double logarithmic plot (bottom).
Blue: \(\chi^2\) distribution \(f_1(\beta)\) with \(d_1=1.51, \beta_0=2.19\),
Green: inverse \(\chi^2\) distribution \(f_2(\beta)\) with \(d_2=0.45,\beta_0=2.19\),
Red: lognormal distribution \(f_3(\beta)\) with \(s=0.87, \mu=0.45\),. }
    \label{fig2}
\end{figure}
It can be seen in Fig.~\ref{fig2} that lognormal, $\chi^2$-
and inverse $\chi^2$ superstatistics all yield a more or less decent fit, though
inverse $\chi^2$-superstatistics seems less favorable.

Still for consistency we also need to check
the validity of Eq.~(\ref{eq4}). We thus also compare the original histogram of returns \(u\) with the following integrals where the parameters take the same values as in Fig.\ref{fig2}:
\begin{equation}
p_i(u)=\int \sqrt{\frac{\beta}{2\pi}} \exp{\left(-\frac{1}{2}\beta u^2\right)}f_i(\beta)d\beta
\;\;\;\;i=1,2,3
\end{equation}

\begin{figure}
\includegraphics[scale=0.5]{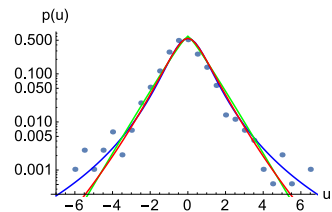}
\caption{Comparison of the histogram of \(u\) (plotted by dots) with the 3 types of superstatistics, integrated with the same
parameters as in Fig.~\ref{fig2}.
Blue: \(\chi^2\) Superstatistics \(p_1(u)\),
Green: inverse \(\chi^2\) Superstatistics \(p_2(u)\),
Red: lognormal Superstatistics \(p_3(u)\). }
\label{fig3}
\end{figure}

As shown in Fig.~\ref{fig3},
for the integrated densities \(\chi^2\) superstatistics seems to fit better to the probability density of \(u\) compared with  lognormal superstatistics and inverse \(\chi^2\) superstatistics.

\begin{figure}
\includegraphics[scale=0.5]{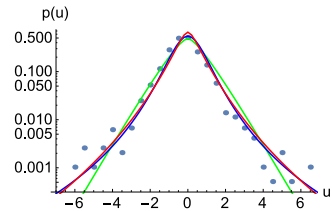}
\caption{Amended Superstatistics
Blue: \(\chi^2\) Superstatistics \(p_1(u)\) with \(d_1=1.51, \beta_0=2.19\),
Green: inverse \(\chi^2\) Superstatistics \(p_2(u)\) with \(d_2=1.2, \beta_0=2.19\),
Red: lognormal Superstatistics \(p_3(u)\) with \(s=1.2, \mu=0.65\). }
\label{fig4}
\end{figure}

Thus, if {\em independent} variation of the volatility parameter in each interval is assumed,
then the data clearly point to  \(\chi^2\) superstatistics, equivalent to Tsallis statistics \cite{tsallisbook}.
On the other hand, independence of $\beta_k$ may not always be a good approximation. There can be strong correlations
of the volatility parameter $\beta_k$, and variations of the time scales where it is
approximately constant. In that case
more complicated dynamics arise,
and one could then possibly get a better fit for the
integrated distributions $p(u)$ if other effective parameters are used. For this reason, we also
     allowed the fitting parameters for \(p_1(u), p_2(u), p_3(u)\) to take on other possible values. The result of this `amended superstatistics' is shown in Fig.~\ref{fig4}.

After the adjustment, we find in Fig.\ref{fig4} that
in fact {\em all three} superstatistics can describe \(p(u)\) quite well.
To distinguish between them, one would need much more data so that the tail behaviour
would be clearer. In practice, more data are available if one considers
price changes on much smaller time scales than days. This will be done in the
next section.

\section{Short time scales}
\begin{figure}
\includegraphics[scale=0.5]{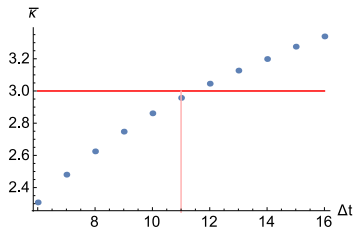}
\caption{Determination of the optimum window size for the 1-minute data
of Alcoa. The intersection with the line \(kurtosis=3\) yields \(T\approx 11\).}
\label{fig5}
\end{figure}

Let us extend our analysis to returns on much smaller time scales.
{A change of statistics as a function of the time scale considered is a common phenomenon
for many complex systems, see e.g. \cite{peng, scafetta} for work
in this direction. Hence it is interesting to also consider
return data on much smaller time scales (say, minutes), and see what is similar and what is different
as compared to the analysis of the previous section.}
Let \(s_i\) be the stock price for every recorded minute,
 in our example chosen as that of Alcoa Inc(AA). The total number of data points is about 1.5 million. We look at the returns
\begin{equation}
 r_i=\log{\left(\frac{s_{i+\tau}}{s_{i}}\right)}
\end{equation}
where \(\tau\) is an integer in units of minutes. The log returns are again normalized
to variance 1:
\begin{equation}
 u_i=\frac{r_i-\left<r\right>}{\sqrt{\left<r^2\right> - \left<r\right>^2}}
\end{equation}
There is one small technical problem for these types of data, as the returns are not
given overnight but only during normal working hours.
          This can lead to big overnight jumps and affect the analysis.
          For this reason, if \(s_{i+\tau}\) and \(s_i\) are from two successive trading days, we removed
          the corresponding \(\log{\left(\frac{s_{i+\tau}}{s_{i}}\right)}\).  \(\tau=1\) means the log return is extracted every minute.
Again we determined the optimal window size, using the same technique as
in the previous section. We obtain $T \approx 11$ (see Fig.~\ref{fig5}).

\begin{figure}
        \includegraphics[scale=0.5]{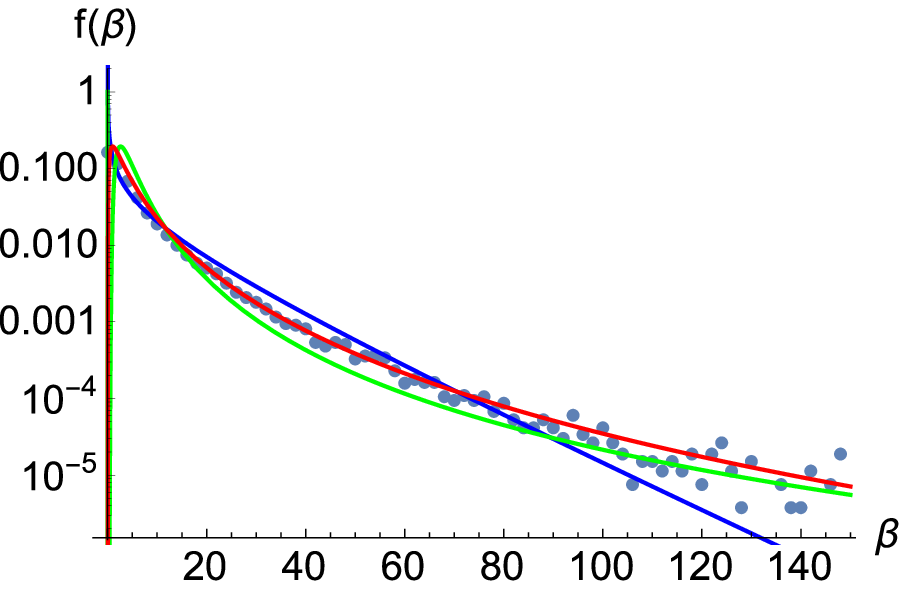}
        \label{fig6a}
        \includegraphics[scale=0.5]{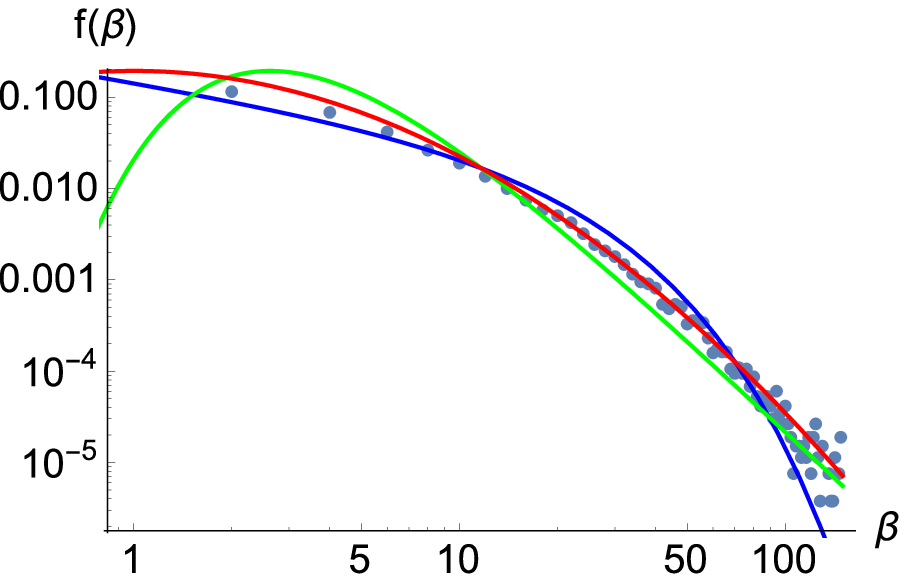}
        \label{fig6b}
    \caption{Best fits that can be achieved for the distribution of the
short-scale volatility parameter \(\beta\) (time scale of returns: 1 minute).
Blue: \(\chi^2\) distribution \(f_1(\beta)\) with \(d_1=0.13, \beta_0=6.33\),
Green: inverse \(\chi^2\) distribution \(f_2(\beta)\) with \(d_2=2.83,\beta_0=6.33\),
Red: lognormal distribution \(f_3(\beta)\) with \(s=1.11, \mu=1.23\), top: log-linear plot, bottom: double
logarithmic plot. }
    \label{fig6}
\end{figure}

\begin{figure}
\includegraphics[scale=0.5]{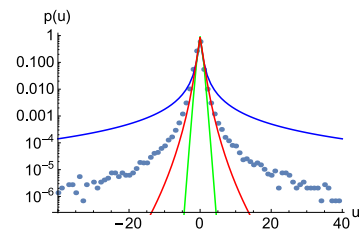}
\caption{Comparison of histogram of \(u\) (plotted by dots) with the integrated
 superstatistics distributions, using the same parameters as in Fig.\ref{fig6}.
Blue: \(\chi^2\) Superstatistics \(p_1(u)\),
Green: inverse \(\chi^2\) Superstatistics \(p_2(u)\),
Red: lognormal Superstatistics \(p_3(u)\). None of the curves is a good fit, indicating the presence of
strong correlations for the volatility parameter $\beta_k$. }
\label{fig7}
\end{figure}

\begin{figure}
\includegraphics[scale=0.5]{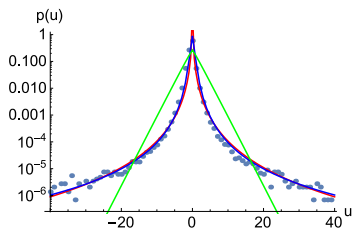}
\caption{Amended Superstatistics
Blue: \(\chi^2\) Superstatistics \(p_1(u)\) with \(d_1=0.36, \beta_0=6.33\),
Red: lognormal Superstatistics \(p_3(u)\) with \(s=2.7, \mu=3.9\), Green: inverse $\chi^2$-superstatistics
($d_2=0.2,\beta_0=1.8$). }
\label{fig8}
\end{figure}

Again this time scale of about 11 minutes makes sense. It is a typical time scale
on which new relevant information becomes available to the traders, leading to
changes in the small-scale volatility.
It also coincides with typical time scales on which observed correlations in
short-term returns start to decay \cite{vicente}.
Our results of fitting the three types of superstatistics are shown in Fig.~\ref{fig6}-\ref{fig8}.

As can be seen in Fig.~\ref{fig6}, the lognormal distribution is by far best fit of $f(\beta)$ if the time scale is 1 minute.

Fig.~7 shows a clear difference as compared to the daily data in Fig.~3: The integrated formula now does {\em not}
give good fits to $p(u)$. The reason is that the $\beta_k$ on a time scale of minutes are not anymore
statistically independent, hence random sampling of Gaussians with different variance is not appropriate anymore.

After the free adjustment in the parameters of \(p_1(u), p_2(u),p_3(u)\), again both \(\chi^2\) and lognormal superstatistics
can provide good fits of $p(u)$. See Fig.~\ref{fig8}.

If one does not allow for parameter amendments, then we can conclude that there is a transition from \(\chi^2\) to lognormal superstatistics when the time scale changes from 1 day to 1 minute. Also, a more general conclusion seems to be that the assumption of a sequence of independent volatility parameters $\beta_k$ is not valid, as we are getting in general differences between the optimum fit of $f(\beta)$ and the corresponding fit of $p(u)$ written as an integral over Gaussians with the same corresponding parameters.

\section{Correlation functions}

\begin{figure}
\includegraphics[scale=0.5]{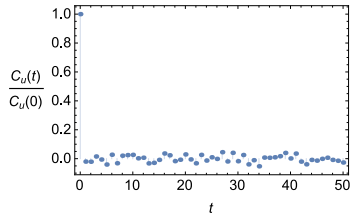}
\caption{Correlation function of log-returns $u$ on a daily time scale for AA shares.
The time unit of $t$ is days.}
\label{fig9}
\end{figure}

\begin{figure}
\includegraphics[scale=0.5]{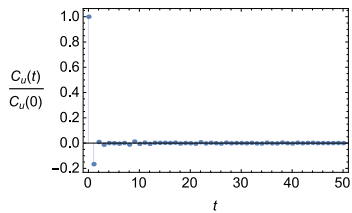}
\caption{Correlation function of log-returns $u$ on a time scale of minutes.
The time unit of $t$ is minutes.}
\label{fig10}
\end{figure}

\begin{figure}
\includegraphics[scale=0.5]{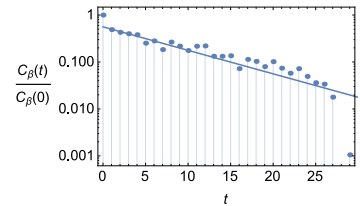}
\caption{Correlation function of volatility $\beta$ for returns on a daily time scale for AA shares.}
\label{fig11}
\end{figure}

\begin{figure}
\includegraphics[scale=0.5]{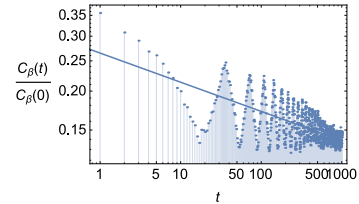}
\caption{Correlation function of volatility $\beta$ for returns on a time scale of minutes.}
\label{fig12}
\end{figure}

For the development of a suitable dynamical model, it is very important to look not
only at probability densities but also on correlation functions and memory effects \cite{cotter}--\cite{evans}. In our case there are two types
of correlation functions: the one of the original data $u_i$,
\begin{equation}
C_u(t) =\frac{1}{N-t} \sum_{i=1}^{N-t} u_i u_{i+t} -\langle u_i \rangle^2
\end{equation}
and those of the volatility parameter $\beta_k$,
\begin{equation}
C_\beta (t) = \frac{1}{n-t} \sum_{k=1}^{n-t} \beta_k \beta_{k+t} - \langle \beta_k \rangle^2.
\end{equation}
Figs.~\ref{fig9}-\ref{fig12} show $C_u(t)/C_u(0)$ and $C_\beta(t)/C_\beta(0)$, both for the daily
returns as well as for the 1-minute returns. As is illustrated in Fig.~9 and 10, $C_u(t)$ decays almost
immediately to zero, both for the daily and minute data.

More interesting is the
correlation function $C_\beta(t)$.
We did an analysis of the decay rates of correlation functions of the volatility for many different
shares from different sectors, the results are summarized in Tab.~1.
We observe that the correlation functions of volatility decay in an exponential way for daily returns,
$C_\beta (t) \sim e^{-\gamma t}$,
whereas for minute return there is a power law decay
$C_\beta (t) \sim t^{-\alpha}$
with a periodic modulation, see Figs.~\ref{fig11}-\ref{fig12} for the example of AA shares.
The strongest correlation decay (largest $\gamma$) on the daily scale is observed for shares from basic materials,
whereas the power law decay (exponent $\alpha$) on the small time scale is largest for healthcare shares and
shares from the consumer
good sector. Note that a strong decay of the volatility correlation function in a sense measures a `volatility of a volatility' and is an interesting quantity to study.
The period of oscillations
that we observe in figures such as Fig.~12 corresponds (roughly) to one trading day and
is consistent with periodic oscillations of intraday
volatility reported previously in \cite{bolleslev}.

\begin{figure}
\includegraphics[scale=0.35]{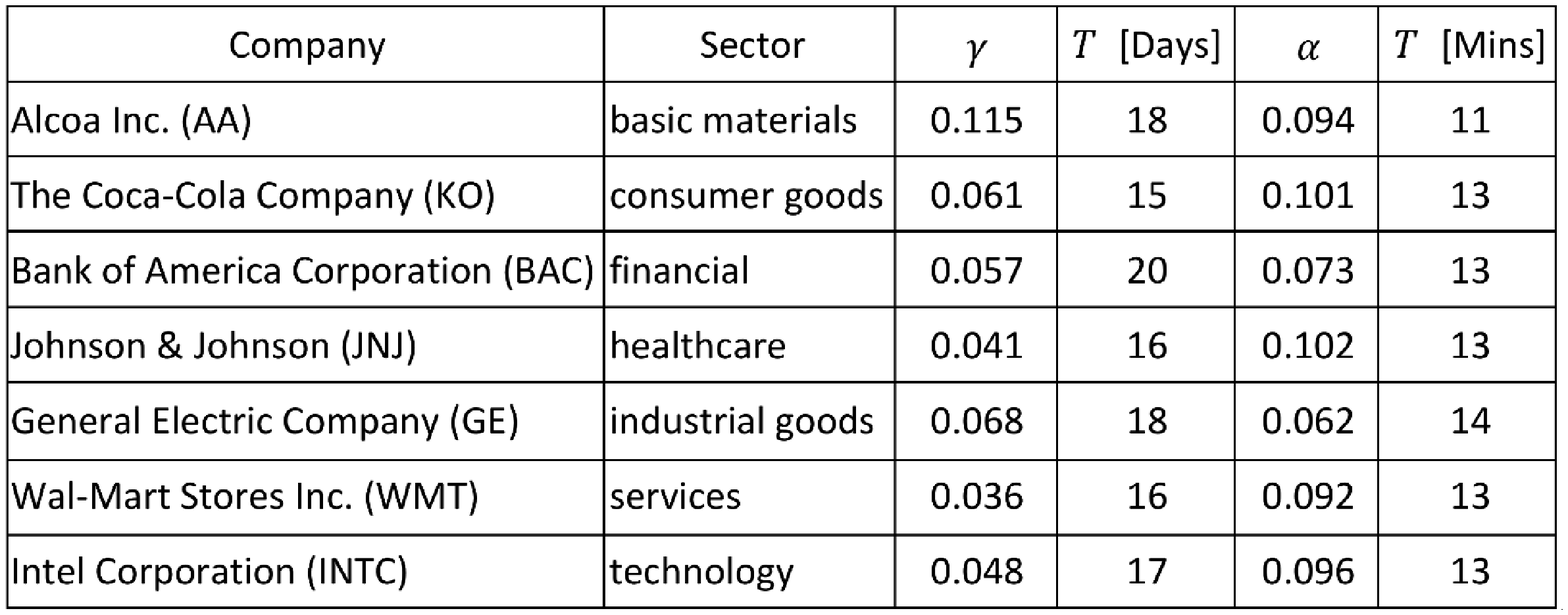}

$\,$

Tab.~1: Decay rates of correlation functions of volatility for shares of different sectors
\end{figure}

\section{Synthetic model}

Based on the results of the previous sections, it is desirable to construct
a simple superstatistical dynamical model that incorporates the
possibility of both lognormal and $\chi^2$ superstatistics on different scales,
and allows for different decay patterns of correlation functions.

Here we propose the following model. We start from a linear superstatistical Langevin equation
\begin{equation}
\dot{u}=-\gamma u +\sigma L(t) \label{pool}
\end{equation}
where $L(t)$ is Gaussian white noise and the `inverse temperature' $\beta$, in
accordance with Einstein's theory of Brownian motion, is
defined as
\begin{equation}
\beta = \frac{\gamma}{2 \sigma^2}.
\end{equation}
{Given a fixed $\beta$, the variance $\langle u^2 (t) \rangle$ is given by
$\frac{1}{2} \langle u^2 \rangle= \beta^{-1}$ for time $t \to \infty$. However,
for the superstatistical version we allow for
fluctuations in the parameter $\beta$.
Then the above Langevin equation is --by construction-- superstatistical as we do not keep the
parameter $\beta$ constant but regard it as a random variable that fluctuates on a large
time scale. Eq.~(\ref{pool}) generates a stochastic process and in the end, after $\beta$ has taken on
many different values, one may rescale the entire time series
$u(t)$ to variance 1 using the variance of the complete time series, as we did in eq.(2) and (18)).}

Let us now consider $n+1$ Gaussian random variables $X_i$, $i=0,1,2, \ldots ,n$
which are statistically independent and have the same variance and mean 0
(except for $X_0$ which may have potentially a different variance and different mean). We then write $\beta$ as
\begin{equation}
\beta =\kappa e^{X_0} +(1-\kappa) (X_1^2+X_2^2+\ldots +X_n^2),
\end{equation}
where $\kappa \in [0,1]$ is a parameter. We now see that if $\kappa =1$, this system
generates lognormal superstatistics, as $\log \beta =X_0$ is a Gaussian random variable.
On the other hand, if $\kappa =0$ this system generates $\chi^2$-superstatistics with
$n$ degrees of freedom, as in this case $\beta =\sum_{i=1}^n X_i^2$ is $\chi^2$
distributed. Choosing any value of $\kappa \in [0,1]$ one can interpolate
between lognormal and $\chi^2$ superstatistics, getting a mixed type of behaviour.

The Gaussian random variables $X_i$ can again be simulated by ordinary
linear Langevin equations of the form
\begin{equation}
\dot{X}_i= - \Gamma X_i + \Sigma L_i (t), \;\;\;\; i=0,\ldots , n
\end{equation}
For constant $\Gamma$ and $\Sigma$ these equations generate the Ornstein Uhlenbeck process,
i.e. a Gaussian Markov process with exponential decay of correlation functions.
More complicated dynamics, leading e.g. to power law decay of correlation functions,
can be constructed if the driving forces in these linear stochastic differential equations
are not Gaussian white noise but more complicated correlated processes,
or critical maps with a near-vanishing Liapunov exponent \cite{tirnakli}.

\begin{figure}
\includegraphics[scale=0.5]{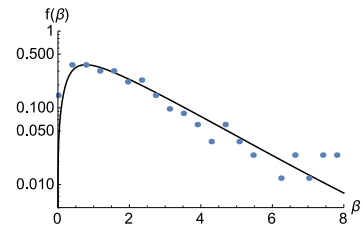}
\caption{Mixed distribution fit to $f(\beta)$ with $\kappa = 0.36$ on the daily time scale.}
\label{fig13}
\end{figure}

\begin{figure}
\includegraphics[scale=0.5]{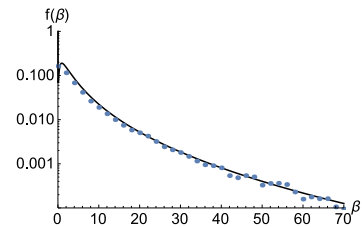}
\caption{Mixed distribution fit to $f(\beta)$ with $\kappa = 0.92$ on the time scale of minutes.}
\label{fig14}
\end{figure}

Fig.~\ref{fig13} and Fig.~\ref{fig14} show that indeed the observed distributions of $f(\beta)$
for Alcoa shares are best fitted by intermediate distributions (a superposition of a lognormal
and $\chi^2$ distribution with appropriate weights). The parameter $\kappa$ increases if one
goes from larger to smaller time scales of returns. The mixed synthetic model is
able to reproduce the transition scenario of observed densities from $\chi^2$ superstatistics
to lognormal superstatistics in a quantitatively correct way, giving good fits on
any time scale.

\begin{figure}
\includegraphics[scale=0.5]{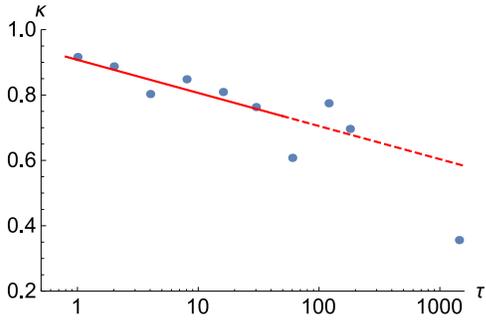}
\caption{Parameter $\kappa$ describing the relative weight of lognormal and $\chi^2$ superstatistics
in the mixed model as a function of the time scale $\tau$ of return. $\kappa$ decreases
if the time scale $\tau$ is increased. For not too big time scales $\tau$ a logarithmic dependence is observed:
The straight line corresponds to a fit of the first six data points of the form $\kappa =0.907 -0.044 \log \tau$.}
\label{fig15}
\end{figure}

 We did this analysis for a variety of time scales $\tau$ of returns, taking again the example of Alcoa shares. In
Fig.~\ref{fig15} we show how the parameter $\kappa$ depends on the time scale of returns.
As expected, the parameter $\kappa$ that best fits the
observations decreases as a function of time scale. In fact we observe a logarithmic dependence if the time
scale is not too big, see the straight line fit in Fig.~15.

{One final remark is at order: One may generalize the superstatistics concept to
more complicated local processes that are not locally Gaussian. Indeed, due to correlations present
on small time scales, and/or due to a lack of clear time scale separation different distributions than Gaussians may
locally be present. In this case one can still superimpose these local distributions by letting a suitable variance parameter
fluctuate. It is remarkable, however, that for the financial data analysed here this generalization to
more complicated non-Gaussian local processes is not necessary: The simplest superstatistics model based on local Gaussians fits our data well,
assuming the interpolating model eq.(23) where the probability distribution of $\beta$ changes as a function of scale.}
\section{Conclusion}

Many investigations of complex systems in the past have focused on the application of a particular
statistics, for example $q$-statistics \cite{tsallisbook}, and then studying the effect of varying system parameters,
 which may change the entropic
index $q$.
Here we have shown that for financial time series  it is sometimes useful
to consider broader classes of statistics and
even proceed from one class of superstatistics to another
when the scale or other system parameters under consideration are changed. The example we considered in detail in this paper were share price returns of various companies.
We provided evidence that
there is a transition scenario from lognormal superstatistics to $\chi^2$ superstatistics, with lognormal
superstatistics giving a better fit to the data on small time scales and $\chi^2$ superstatistics
($=q$-statistics) on larger time scales. We constructed a hybrid superstatistical model that allows to
implement both types of superstatistics, with a weighting parameter $\kappa$
that describes how far away we are from one of the two cases. Correlation functions
of the extracted superstatistical volatility parameter $\beta_k$ were shown to exhibit different qualitative behavior as
a function of the time scale of returns, with exponential decay on large time scales and power law decay on small time scales, modulated by intraday periodicity. The decay parameters of the exponential or power law decay
were extracted from the data and were shown to depend slightly
on the sector of shares considered. The general transition scenario from lognormal to $\chi^2$ superstatistics
as a function of the time scale of returns, however,
is a general phenomenon and occurs for all sectors in a similar way.




\end{document}